\documentclass[pra,twocolumn,showpacs,preprintnumbers,amsmath,amssymb,superscriptaddress]{revtex4}

\usepackage{color}
\usepackage{graphicx}
\usepackage{dcolumn}
\usepackage{bm}
\usepackage{amssymb}

\begin{document}


\title{Suppression of ac Stark shift scattering rate due to non-Markovian behavior}

\author{Muzaffar Q. Lone}
\affiliation{National Institute of Informatics, 2-1-2
Hitotsubashi, Chiyoda-ku, Tokyo 101-8430, Japan}
\affiliation{Department of Physics, University of Kashmir, Srinagar-190006, India}
\affiliation{TCMP division, Saha Institute of Nuclear Physics, Kolkatta-700064, India}

\author{Tim Byrnes}
\affiliation{New York University, 1555 Century Ave, Pudong, Shanghai 200122, China}
\affiliation{NYU-ECNU Institute of Physics at NYU Shanghai, East China Normal University, Shanghai 200062, China}
\affiliation{National Institute of Informatics, 2-1-2 Hitotsubashi, Chiyoda-ku, Tokyo 101-8430, Japan}

\date{\today}

\begin{abstract}
The ac Stark shift in the presence of spontaneous decay is typically considered to induce an effective
dephasing with a scattering rate equal to $ \Gamma_s |\Omega|^2/\Delta^2 $, where $ \Gamma_s $ is the spontaneous
decay rate, $ \Omega $ is the laser transition
coupling, and $ \Delta $ is the detuning.  We show that under realistic circumstances this dephasing rate may be strongly 
modifed due to non-Markovian behavior. The non-Markovian behavior arises due to an effective modification of the 
light-atom coupling in the presence of the ac Stark shift laser.  An analytical formula for the non-Markovian 
ac Stark shift induced dephasing is derived.  We obtain that for narrow laser linewidths the effective dephasing rate 
is suppressed by a factor of $ Q^2$, where $ Q $ is the quality factor of the laser.    
\end{abstract}

\pacs{03.75.Dg, 37.25.+k, 03.75.Mn}
\maketitle

{\it Introduction} --- The ac Stark effect, or light shift, is one of the fundamental phenomena in atom-light interactions \cite{meystre01}.  
The basic effect results from applying an off-resonant laser between two atomic levels,
which creates an energy shift equal to $ |\Omega|^2/\Delta $, where $ \Omega $ is the laser 
transition coupling and $ \Delta $ is the energy detuning \cite{autler55}. The effect is 
responsible for a wide variety of techniques in atomic, molecular, and optical physics. 
Since the energy shift is proportional to the intensity of the light, it is the basis of
optical traps, where a spatially varying light field creates regions of potential
minima \cite{chu86}. Optical lattices created by two counterpropagating laser beams in a
standing wave configuration creates a periodic potential due to the same effect \cite{metcalf99}.
Laser cooling techniques such as Sisyphus cooling also rely on the ac Stark effect by exchanging
kinetic energy for potential energy in an optical lattice geometry \cite{dalibard89,metcalf99}. 
Such trapping and cooling techniques are fundamental for quantum information applications 
such as quantum simulation \cite{buluta09} and quantum computation \cite{brennen99} based 
on neutral atoms. The control of how much the light shift occurs is an important problem 
for atomic clocks, where magic wavelengths are used to ensure cancellations of the shifts
\cite{katori03,ye08,rosenbusch09}.   The observation of the ac Stark effect is now of fundamental importance 
to other quantum systems being investigated for quantum information applications, such as 
superconducting qubits \cite{schuster05}, Bose-Einstein condensates \cite{byrnes12,pyrkov13}, and 
quantum dots \cite{unold04}.

\begin{figure}
\centering
\includegraphics[scale=0.28]{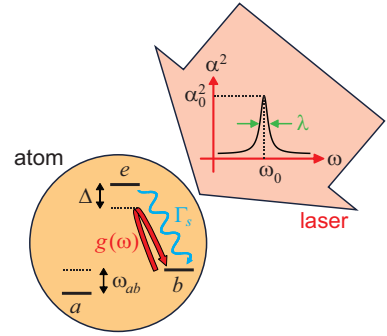}
\caption{\label{ORF} (Color online) The ac Stark shift configuration.  
An atom is illuminated with an off-resonant laser between levels $ b $ and $ e $. 
The laser is assumed to have a Lorentzian distribution with a central frequency $ \omega_0 $, 
maximum intensity $ |\alpha_0|^2 $ and linewidth $ \lambda $. The frequency dependent coupling 
between the levels $ b $ and $ e $ is $ g (\omega) $, and the spontaneous emission rate is $ \Gamma_s $.  
The laser is off-resonant by a frequency $ \Delta $. $ a,b $ label the ground states of the atom.  
The coherence between the ground states is dephased due to the ac Stark shift of the laser.  }
\end{figure}

Under the Markovian approximation the ac Stark shift in the presence 
of spontaneous decay introduces a dephasing (the ``scattering rate'') on the ground states coupled to the laser according to \cite{grimm00} is 
\begin{align}
\Gamma_M = \frac{\Gamma_s |\Omega|^2}{\Delta^2},
\label{standarddephasing}
\end{align}
where  $ \Gamma_s $ is the spontaneous decay rate. It is often assumed that under typical experimental 
situations that the Markovian approximation is sufficient to capture the 
basic effects of decoherence in quantum systems.  However, this assumption 
depends upon several factors, and in particular for atomic systems
(the ``system''), the effects due to radiation field (the ``bath'') are known to be either Markovian
or non-Markovian depending on the circumstances \cite{beuer00,gardiner04}. 
It has been shown that such effects are important to understand the dynamics as shown in several 
experimental systems \cite{bigot91,galland08}, and has been even shown to be controllable such as to enter both regimes \cite{liu11}.
Understanding the nature of non-Markovian dynamics is naturally a very important topic for quantum information 
science, where the aim is to control a quantum system for use in technological applications \cite{terhal05,ban06,wolf08,pastawski11}.
For this reason several works have studied the fundamental problem of a two-level 
system in a variety of non-Markovian settings 
\cite{divincenzo05,maniscalco06,shresta05,haikka10,vacchini10,tong10,apollaro11,haikka12,breuer10}. 
Most works have focused on the modification of spontaneous decay in the non-Markovian regime.  
However, to our knowledge, no calculation of the non-Markovian scattering rate has been performed before. 
In this paper, we analytically calculate the ac Stark shift dephasing in the non-Markovian regime when illuminated by a narrow linewidth laser.

The standard way to calculate the ac Stark shift scattering rate (\ref{standarddephasing}) is by assuming a 
master equation involving a ground and excited state, with the excited state having the possibility of 
spontaneous decay \cite{pyrkov13} (see Fig. \ref{ORF}).  Specifically, the coherent part of the evolution is determined by the Hamiltonian
\begin{align}
{\cal H}/\hbar = \Delta | e \rangle \langle e | + (\Omega |b \rangle \langle e| + \mbox{H.c.})  
\end{align}
in the presence of spontaneous decay described by a master equation
\begin{align}
\frac{d \rho}{dt} = -\frac{i}{\hbar} [ {\cal H} , \rho] + \frac{\Gamma_s}{2} {\cal L} [| b \rangle \langle e |] \rho ,
\label{master}
\end{align}
where the Lindblad decay term is $ {\cal L} [O] \rho = 2 O \rho O^\dagger  - O^\dagger O \rho - \rho O^\dagger O $.
We show that this commonly assumed framework in fact misses several aspects which turn out to be rather important to the problem. 
The first is that the spontaneous
decay rate $ \Gamma_s $ is derived without the presence of the laser.  The laser only enters the problem after the (\ref{master}) 
is constructed.  As we show, if the dephasing is derived in the presence of the laser, 
qualitatively different results are obtained.  The second related issue is that when $ \Gamma_s $ is derived 
in (\ref{master}),
the Markovian assumption is quite reasonable, as the coupling to the environment can
be considered to be smooth.  However, in the presence of a laser, the effective coupling no longer satisfies this, 
and non-Markovian effects become more important. 
We show in this paper that these aspects strongly modify the standard dephasing rate (\ref{standarddephasing}). 
In particular, under typical experimental parameters we expect that non-Markovian effects are in fact very important to
consider.  Remarkably, for narrow linewidth lasers this results in dephasing rates that are greatly suppressed in comparison to the standard result (\ref{standarddephasing}) .

{\it Model Hamiltonian} ---  Our derivation of the ac Stark shift induced dephasing proceeds as follows.  
We consider an excited state $ | e \rangle $ and ground states of an atom $ |a,b \rangle $, 
coupled to a continuum of electromagnetic modes according to 
\begin{align}
\label{cqed}
H/\hbar = & \omega_e | e \rangle \langle e | + \omega_a | a \rangle \langle a |+ \omega_b | b \rangle \langle b |  + \sum_{k} \omega_{k}p^{\dagger}_k p_k \nonumber \\ & +  \sum_{k} \left(  g_k |e \rangle \langle b| p_k + \mbox{H.c.} \right)
\end{align}
where  $p_k,\omega_k$ are the respective annihilation operator and frequency of the $k$th bath mode, $ \omega_{a,b,e} $ are
frequencies of the levels $ a,b,e $, and $g_k$ describes the coupling to the electromagnetic field via electric dipole 
approximation (see Fig. \ref{ORF}). We assume for simplicity that the photons only couple to level $ b $, 
which may arise in practice due to selection rules of the laser transition. 
Our results can be easily generalized as long as the bath modes for levels $ a, b $ may be treated independently.  
In the usual treatment one would eliminate the bath modes in (\ref{cqed}) to obtain above the master equation 
under a Markovian approximation, without the presence of the laser. However, in our procedure instead we will
derive the dephasing master equation in the presence of the laser.  In the presence of the off-resonant laser,
there will be a  second order virtual process of the atom being excited to higher energy level and relaxing back to the ground state. 
For a large detuning $ \Delta $ between the laser frequency $\omega_0$  
and the atomic transition frequency, we can eliminate the excited state adiabatically. Setting the zero point energy at level $ b $ and following the standard adiabatic elimination procedure by setting $ i \frac{d | e' \rangle }{dt} = H | e' \rangle = 0 $, where $ | e' \rangle = e^{i \omega_0 t} | e \rangle $, we obtain 
within the rotating wave approximation
\begin{align}
H_{\mbox{\tiny eff}}/\hbar = & \frac{\omega_{ab}}{2} \sigma^z + \sum_{k} \left( \omega_{k} - 
\frac{|g_k|^2}{\Delta} \right) p^{\dagger}_k p_k \nonumber \\
& +  \sum_{k} \frac{|g_k|^2}{\Delta} \sigma^z  p_k^{\dagger} p_k  
\end{align}
where $\sigma^z = |a \rangle \langle a | - |b \rangle \langle b | $, $ \omega_{ab} = 
\omega_a -\omega_b $, $ \Delta = \omega_0 - (\omega_e -\omega_b) $ and constant terms
have been dropped. This Hamiltonian describes the phase coupling occurring during the virtual 
transition to excited state and relaxing back to the ground state. 

The presence of the laser may be taken into account by displacing the photon operators with the 
Hamiltonian $ H_{\mbox{\tiny laser}} = \sum_{k} F_k p_k + F_k^* p_k^\dagger  $, where $ F_k $ is the laser amplitude. 
The laser displacement may be eliminated by a transformation $ q_k = p_k +\alpha_k = D^{-1}(\alpha_k) p_k D(\alpha_k) $, where $ \alpha_k $
is the resulting coherent state amplitude, 
$ D(\alpha_k) = e^{\alpha_k p^\dagger_k - \alpha_k^* p_k} $ is the displacement operator \cite{scully97}. The Hamiltonian is then 
\begin{eqnarray}
H_{\mbox{\tiny eff}}/\hbar  &=&\frac{ \omega_r}{2} \sigma^z + \sum_{k} \omega_{k} q^{\dagger}_k q_k  - \sum_{k} \frac{|g_k|^2 }{\Delta} \sigma^z  (\alpha_k q_k^{\dagger}+ \mbox{H.c.}) \nonumber  \\
\end{eqnarray}
where $\omega_r = \omega_{ab} + \sum_k \frac{|g_k \alpha_k |^2}{\Delta}$ is the the renormalized frequency between the ground states, and we have assumed that $ \omega_{k} \gg \frac{|g_k|^2}{\Delta} $ as the bath modes that are relevant to the atomic transition are much larger than the second order shifts in the energy.  Moving to interaction picture with respect to bath and the system,
the interaction Hamiltonian can be written as
\begin{eqnarray}
 H_I(t)/\hbar = -\sum_{k} \frac{|g_k|^2 }{\Delta} \sigma^z  (\alpha_k q_k^{\dagger} e^{i \omega_k t}+ \mbox{H.c.}). 
\end{eqnarray}

{\it Non-Markovian scattering rate} --- The time evolution of the bath and spin can be obtained by evolving the total density matrix $\rho_T(0)$ with 
the time evolution operator $U(t)$ from an initially unentangled state 
$\rho_{s}(0) \otimes \prod_k  |\alpha_k \rangle_k \langle \alpha_k|_k $ in terms of the original variables $ (\sigma, p_k ) $.  For the displaced basis $ (\sigma, q_k ) $, the initial state is 
\begin{equation}
\rho_{T} (0) = \rho_{s}(0) \otimes \prod_k  |0 \rangle_k \langle 0 |_k . 
\end{equation}
In the interaction picture, the time evolution operator $U(t)$ is given by
\begin{eqnarray}
U(t) = T_{\leftarrow} \exp[-\frac{i}{\hbar} \int_0^{t}ds H_I(s) ]
\end{eqnarray}
where $ T_{\leftarrow}$ is the time ordering operator. The above time ordered product is usually difficult to evaluate.
However, the interaction Hamiltonian $H_I$ in our case simplifies its evaluation. We note that
$[\sigma^z(t),H_I(t')]=0$ and the unequal time commutator
\begin{eqnarray}
[ H_I(t), H_I(t')] = -2i \hbar^2 \Phi(t-t'),
\end{eqnarray}
which is constant in all operators as $ \Phi(t) = \sum_k \frac{|g_k|^4 |\alpha_k|^2}{\Delta^2} \sin\omega_k (t)  $.   
This allows us to write the time ordered evolution operator in the form
\begin{align}
U(t) & =   \exp \left[ i \int_0^t dt_1 \int_0^{t_1} dt_2 \Phi (t_1-t_2) \Theta(t_1-t_2)   \right]  \nonumber \\
 & \times \exp  \left[ - \sigma^z \sum_k \frac{| g_k|^2 }{\Delta} ( \alpha_k f_k(t)q^{\dagger}_k - \mbox{H.c} ) \right]
\label{phase}
\end{align}
where $f_k(t)= \frac{e^{i\omega_kt} -1}{\omega_k}$. The first factor in 
$U(t)$ is a
global phase factor and does not involve any dynamical change to the time evolution of the system. 

We may now obtain the effective dynamics of the spin alone at a time $ t $ by taking the trace
over bath degrees of freedom $ \rho_{s}(t) = \mbox{Tr}_B [ U(t)  \rho_{T}(0)  U^{\dagger}  (t)]  $.
Taking $ \rho_s^{\sigma \sigma'} = \langle \sigma |\rho_s |\sigma' \rangle  $ with $ \sigma^z| \sigma \rangle =\sigma | \sigma  \rangle $, we obtain
\begin{align}
\rho_s^{\sigma \sigma'} (t)=  \rho_s^{\sigma \sigma'} (0) \exp [- \frac{(\sigma -\sigma')^2}{4} \Gamma(t)] 
\label{rho}
\end{align}
where 
\begin{align}
\Gamma(t) = \sum_k \frac{4 |g_k|^4 |\alpha_k|^2}{\Delta^2} \frac{(1-\cos\omega_k t)}{\omega_k^2}  
\label{decfunc}
\end{align}
As expected the dephasing preserves the diagonal elements of 
$\rho_s(t)$ and eventually reduces off-diagonal density matrix elements to zero. 

In most experimental situations the ac Stark shift is induced by a narrow linewidth laser, 
in which case the coherent field distribution is well described by a Lorentzian
\begin{eqnarray}
\alpha^2(\omega)
=  \frac{|\alpha_0|^2}{\pi } \frac{\lambda^2}{(\omega-\omega_0)^2 + \lambda^2} . 
\end{eqnarray}
Here $\lambda$ represents the spectral width, $ |\alpha_0|^2 $ is the intensity, and $ \omega_0 $ is 
the center frequency of the laser.  The decoherence function (\ref{decfunc}) may be evaluated by using the 
standard electric dipole approximation in free space to obtain an expression for $ g_k $ as shown 
in the Appendix. The decoherence function then reduces to  
\begin{align}
\Gamma(t) = & \frac{\Gamma_s |\Omega|^2}{\Delta^2 } \Big[ \frac{\lambda^2 t }{\omega_0^2 +\lambda^2 } \nonumber \\
& + \frac{\lambda [(\omega_0^2-\lambda^2)(1- e^{-\lambda t}\cos \omega_0 t) 
- 2 \omega_0 \lambda e^{-\lambda t} \sin \omega_0 t]}{2(\omega_0^2+\lambda^2)^2}  \Big]. 
\label{gammafunc}
\end{align}
The coefficient of  (\ref{gammafunc})
is precisely that given in the standard expression for the scattering rate $\Gamma_M $ in Eq. (\ref{standarddephasing}).  In terms of the dimensionless time $ \tau = \Gamma_M t $, there are two non-trivial parameters in 
(\ref{gammafunc}): the laser quality factor $ Q = \omega_0 / \lambda $, and the ratio $ R = \lambda/\Gamma_M $ which allows us to rewrite the decoherence function as
\begin{align}
\Gamma(\tau) & =  \frac{\tau}{Q^2+1} \nonumber \\
& + \frac{(Q^2-1)(1- e^{-R \tau }\cos (QR \tau)) - 2 Q e^{-R \tau} \sin (QR \tau)}{2R (Q^2+1)^2} .
\label{dimlessgamma}
\end{align}

\begin{figure}
 \centering
 \includegraphics[scale=0.33]{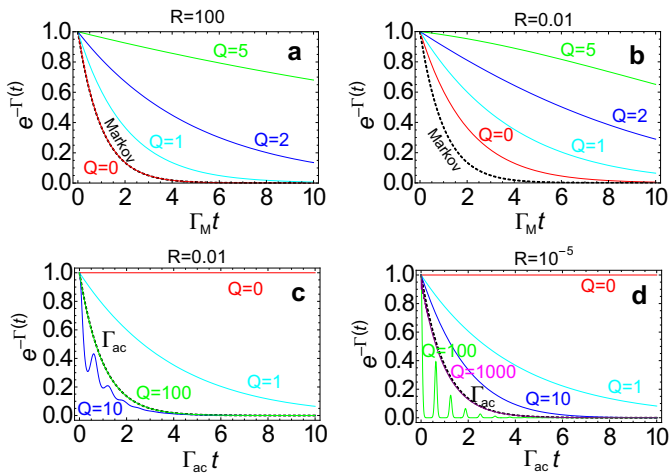} 
 \caption{(Color online) Decay of coherence with time at different values of the parameter $ Q = \omega_0/\lambda $ and $ R = \lambda/ \Gamma_M $. Plots are for $ Q $ values as marked with (a) $ R = 100 $, (b) (c)  $ R =0.01 $,  (d) $ R =10^{-5} $.  Plots (a) and (b) are in units of the Markovian decay time $ \Gamma_M = \frac{\Gamma_s \Omega^2}{\Delta^2} $. Dotted line shows the Markovian approximation 
$ e^{- \Gamma_M t} $. Plots (c) and (d) are in units of the ac Stark shift dephasing rate $ \Gamma_{ac} = \Gamma_M/Q^2 $.  The dotted line shows the approximation $ e^{- \Gamma_{ac} t} $. }
 \label{Plot}
 \end{figure}

{\it Analysis of decoherence function} --- Let us first understand how (\ref{dimlessgamma}) reduces to the standard Markovian limit.  
Figure \ref{Plot}(a) shows the coherence factor $ e^{-\Gamma(t)} $ for various values of $ Q $ with $ R \gg 1 $.  
We see that the the Markovian curve is recovered for $ Q=0 $ and $ R \gg 1 $.  Physically, this can be understood 
by looking at the role of the various parameters.  The  typical time scale over which there exists an intrinsic 
time evolution of the system is $ \sim 1/\omega_0 $. Meanwhile as $\lambda$ is the spectral width of the coupling, 
the bath correlation time $ \sim 1/ \lambda $. The Markovian approximation implies that system evolves over large 
time compared to very fast bath dynamics, implying $Q \ll 1$ and $ R \gg 1 $. In this limit (\ref{dimlessgamma}) 
reduces to $ \Gamma(t) = \Gamma_M t $, agreeing with standard Markovian dephasing (\ref{standarddephasing}).

However, the Markovian limit is not always the physically relevant regime. In practice the linewidth $ \lambda $ of the laser is much narrower than the transition frequency $ \omega_0 $, implying $ Q \gg 1 $. 
Taking numbers for $ ^{87} \mbox{Rb} $ for example \cite{DS}, the spontaneous emission rate is $ \Gamma_s \approx 19 $ MHz, which may typically give $ \Gamma_M \sim $ MHz.  The transition frequency is $ \omega_0 \approx 300 $ THz, and the linewidth in the range $ \lambda \approx $ kHz to MHz, implying $ R < 1 $.  These parameters suggest that the Markovian limit is in fact the {\it opposite} regime in terms of the magnitudes of $ Q $ and $ R $ to the physically relevant regime.  

The decay of the off-diagonal terms of the density matrix at various values of $Q $ for the more physically relevant $ R < 1 $ regime is shown in Fig. \ref{Plot}(b). It can be observed that as $Q$ increases, there is a larger deviation from Markovian behavior. The general trend is similar for all values of $ R $ as can be seen on comparison to Fig. \ref{Plot}(a). The term that is primarily responsible for this is the first term in (\ref{dimlessgamma}), where the $ Q^2 + 1 $ term appears in the denominator.  Considering that $ Q \gg 1 $ is the physically relevant regime, this suggests defining another timescale
\begin{align}
\Gamma_{ac} = \frac{\Gamma_M}{Q^2} = \frac{\Gamma_s |\Omega|^2 \lambda^2}{\Delta^2 \omega_0^2} .
\label{gammaac}
\end{align}
Figure \ref{Plot}(c)(d) shows the behavior of the coherence function in terms of the rescaled time $ 
\tau' = \Gamma_{ac} t $ for two different values of $ R$.   We see that for large $ Q $ the 
coherence function coincides with exponential decay in units of the rescaled time.

Before reaching the limiting behavior of exponential decay with rate (\ref{gammaac}), we observe some oscillatory behavior Figure \ref{Plot}(c)(d). This occurs clearly due to the second term in (\ref{dimlessgamma}).  For $ Q \gg 1 $ we may approximate this as
\begin{align}
\Gamma(\tau') & \approx  \tau'  + \frac{(1- e^{-R Q^2 \tau' }\cos ( R Q^3  \tau'))}{2R Q^2} .
\label{approxgamma}
\end{align}
In order for the oscillatory term to be visible, we require the exponential decay term in 
(\ref{approxgamma}) to be not too fast giving $ R Q^2 < 1 $, but simultaneously the oscillation 
frequency should be faster than the overall decay envelope $ R Q^3 > R Q^2 > 1 $.  The strongest 
oscillations therefore occur when $ R Q^2 \sim 1 $, which agree with the numerical plots in Figure 
\ref{Plot}(c)(d). For larger values of $Q$, the oscillatory term in (\ref{approxgamma}) is suppressed, and 
we return to pure exponential decay with a rate (\ref{gammaac}). For typical experimental parameters as given above
we have $ R Q^2 \gg 1 $, hence much lower values of $ Q $ are necessary to see the oscillatory behavior.  The deviation from an exponential decay can be attributed to the memory effects 
developed initially, typical of non-Markovian behavior. As $ R Q^2 = \omega_0^2/ \Gamma_M \lambda $ it is natural to expect that such a parameter should be sufficiently large to observe oscillations, which may be achieved by having a narrow linewidth and low spontaneous emission rate. 
In general, three time scales in an open system exist to characterize non-Markovian dynamics: (i) the time scale of the system $ \sim 1/\omega_0 $ (ii) the time scale of
the bath $ \sim 1/\lambda $ (iii) the mutual time scale arising from the coupling between the system and the bath $ \sim 1/\Gamma_M $ \cite{zhang12,haikka10,cirone09}. In this case the criterion for the strongest oscillatory behavior are satisfied when these are all approximately the same.

{\it Comparison to past works} --- Several works have studied non-Markovian dynamics of a qubit with a Lorentzian bath before \cite{maniscalco06,vacchini10,haikka10}. 
Here we discuss whether the results of these past works could be used to derive our primary result (\ref{dimlessgamma}). 
One potential approach to deriving this equation would be to use the standard expression for the ac Stark shift scattering rate (\ref{standarddephasing}), and replace the non-Markovian spontaneous decay in the place for $ \Gamma_s $.
We find that such an approach does not yield the same results as our (\ref{dimlessgamma}). 

Taking the example of Ref. \cite{vacchini10}, there the non-Markovian decay of the density matrix $ \rho $ of a two-level system in a Lorentzian bath is derived.  From Ref. \cite{vacchini10}, the excited state density matrix element obeys
\begin{align}
\rho_{ee}(t) =e^{-\lambda t}\left[ \cosh(\frac{\lambda t}{2}\delta)+ \frac{1}{\delta}\sinh(\frac{\lambda t }{2}\delta) \right]^2 \rho_{ee}(0), \label{prevdecay}
\end{align}
where $\delta = \sqrt{1- \frac{2 \Gamma_s}{\lambda}}$ and $\Gamma_s$ is the Markovian spontaneous decay 
rate. In the weak coupling limit $ \Gamma_s \ll \lambda $, there is a very broad coupling to many frequency modes,
which gives Markovian behavior.  Here $ \delta \approx 1 - \frac{\Gamma_s}{\lambda} $ and the decay function 
(\ref{prevdecay}) gives purely exponential behavior.  To first order the decay function may be approximated as
$ \rho_{ee}(t) \approx e^{-\Gamma_s t/2} \rho_{ee}(0) $, which is nothing but standard Markovian spontaneous decay.  In this limit the ac Stark scattering rate would be exactly 
the same as the standard expression (\ref{standarddephasing}). Meanwhile, in the strong coupling limit $ \Gamma_s \gg \lambda $, the linewidth of the laser is extremely narrow, which gives rise to strongly non-Markovian behavior.  Here we may approximate 
$\delta =i\sqrt{\frac{2 \Gamma_s}{\lambda}} $ 
and 
\begin{align}
\rho_{ee}(t) = e^{-\lambda t}\left[
\cos \left(  \Omega_{NM} t \right)+ \sqrt{\frac{\lambda}{2 \Gamma_s}}\sin \left( \Omega_{NM} t \right) 
\right]^2  \rho_{ee}(0) \nonumber
\end{align}
which corresponds to damped oscillations at frequency $ \Omega_{NM}= \sqrt{\lambda \Gamma_s/2}$ and a decay 
envelope with rate $\lambda$. Taking $ \Gamma_s = \lambda $ and substituting into (\ref{standarddephasing}) we would obtain 
$\Gamma_{ac}=\frac{\lambda |\Omega|^2}{\Delta^2}$.  This again does not yield the scaling factor $Q^2$ as 
derived in (\ref{gammaac}). 

The above approach and ours mainly differ in what stage the bath is eliminated to produce an effective master equation.  It is helpful to examine what order effect we are examining in terms of the original couplings $ g_k $.  In our derivation we start with an effective coupling of the form 
$ \propto g_k^2 \sigma^z p^\dagger_k p_k $ to the bath modes which directly gives the dephasing of the qubit.  As is typical of all master equation derivations, the dephasing rate is the square of the coupling, hence our final expression is effectively an expression that is fourth order in 
$ \Gamma_{ac} \propto g_k^4 $, as seen in  (\ref{decfunc}).  On the other hand, if one uses the spontaneous emission as a starting point, the dephasing is not calculated directly, but is inferred from the non-Markovian spontaneous emission 
rate.  In this case the bath is eliminated at first order in the coupling, resulting in the spontaneous emission being a second order effect $ \Gamma_s \propto g_k^2$.  Then if we substitute this into (\ref{standarddephasing}), there is another factor of $ g_k^2$  from the $\Omega^2$, hence both give expressions that are fourth order in $ g_k$.  However, our derivation is a direct fourth order calculation, whereas (\ref{standarddephasing}) is combining two second order results.  Combining two second order results does not yield generally the same results as a direct fourth order calculation, which is the case here as well.  

{\it Conclusions} ---  We have derived the ac Stark shift scattering rate in the fully non-Markovian regime. 
In contrast to the standard procedure where the laser field is put in after the bath is eliminated,
in our procedure the bath is eliminated only at the end to obtain the dephasing rate. 
While our results reduce to the standard expression for the scattering rate in the Markovian limit,
we find that this is not the typical physically relevant regime.  
In the physically relevant regime, the scattering rate is better approximated 
by $ \Gamma_{ac} $ (Eq. (\ref{gammaac})), which is equal to the standard Markovian 
scattering rate divided by the square of the laser quality factor $ Q $, which can be quite large in practice.
We note that in terms of the order of the result, both are fourth order in terms of the atomic transitions.
Thus the difference does not result from a difference in order of the calculations, and is a result of a consistent non-Markovian 
treatment of the dephasing induced in the presence of the laser. 
Furthermore, (\ref{gammaac}) cannot be derived by simple methods using past works deriving non-Markovian spontaneous emission in a Lorentzian bath \cite{maniscalco06,vacchini10,haikka10}. 
We attribute this to the fact that our result is a direct fourth order calculation, whereas (\ref{standarddephasing}) combines two second order effects.   

Physically, we may understand the difference as follows.  In the standard approach,
dephasing may occur because during one of the transition of the ac Stark shift, 
instead of being driven back down by the laser, spontaneous emission may occur.  
As the spontaneous emission in 
(\ref{master})  is derived without the presence of the laser, 
the excited mode couples to the full range of frequencies of the bath modes.  On the other hand,
in our derivation, the presence of the laser linewidth bosonically enhances emission into the bath
modes that are excited by the laser.  Thus spontaneous emission occurs only at those modes that are
excited by the laser, which is much narrower than that predicted by 
(\ref{master}). 
For fewer bath modes, the dephasing rate is naturally much smaller, in this case by a factor 
$ Q^2 $. 

In most practical applications of the ac Stark shift, 
the experimental parameters are chosen such that even assuming the 
rate (\ref{standarddephasing}), the detuning is made large enough that 
the scattering rate is made negligible compared to the light shift.
The current result would imply that particularly for narrow linewidth lasers, 
the detuning requirements could be made less, offering an alternative approach 
to controlling the dephasing.  This would be important for applications where spontaneous emission
is a serious drawback of using excited states, such as for quantum information processors, 
quantum simulators, and quantum metrological applications.

{\it Acknowledgments} --- MQL acknowledges the support and hospitality provided by NII under a non-MOU grant, 
and to the University of Kashmir for allowing a visiting fellowship. This work is 
supported by the Shanghai Research Challenge Fund, Transdisciplinary Research Integration Center, 
the Inamori Foundation, NTT Basic Laboratories.

{\it Appendix} ---  In order to evaluate (\ref{decfunc}), 
we will write the coupling $g_k$ in the electric dipole approximation  \cite{walls08,anastopoulos00} as 
$
g_k = - i \sqrt{\frac{\omega_0^2}{2 \hbar \omega_k \epsilon_0}} \bm{A}_k( \bm{r}) \cdot \bm{d}
$
%
where $\bm{A}_k( \bm{r})$  defines the vector potential in electric dipole approximation, and  $ \bm{d}$ is the dipole operator for the transition. 
In terms of vacuum Rabi frequency 
$
g_0 =2 i \sqrt{\frac{\omega_0}{2 \hbar \epsilon_0}} \bm{A}_k( \bm{r}) \cdot  \bm{d} ,
$
%
we can write the coupling as 
$
g_k = -\frac{g_0}{2} \sqrt{\frac{\omega_0}{\omega_k}} .
\label{gk}
$
%
Changing the summation in equation (\ref{decfunc}) to integration over $k$, and using the form of $g_k$ we write
\begin{eqnarray}
\!\!\!\!\!\!\!
\Gamma(t) 
= \frac{V\omega_0^2}{16 \pi^3 \Delta^2}   \int \!\!dO~ |g_0|^4 \int \!\!dk ~ k^2 |\alpha_k|^2 \frac{1- \cos \omega_k t}{\omega_k^4}  \nonumber
\label{decfunck}
\end{eqnarray}
%
Integrating over solid angle $dO$, we write 
$
 \int\!\! dO~ |g_0|^4 = \frac{16 \pi \omega_0^2 |d|^4}{5 \hbar^2 \epsilon_0^2 V^2}
$
%
where $d$ is the dipole moment of the transition. Now changing the variables of integration in the above equation 
to $\omega$ using the dispersion relation for free space $\omega= ck$, and also using the form of 
 $\alpha(\omega)$ the equation for the decoherence function $\Gamma(t)$ simplifies after integrating using method of residues:
\begin{align}
\Gamma(t) = & \frac{\omega_0^4 |\alpha_0|^2 |d|^4 \lambda}{5 \pi^3 \epsilon_0^2 \hbar^2 c^3 \Delta^2 } \Big[ \frac{\lambda t }{\omega_0^2 +\lambda^2 } \nonumber \\
& + \frac{(\omega_0^2-\lambda^2)(1- e^{-\lambda t}\cos \omega_0 t) 
- 2 \omega_0 \lambda e^{-\lambda t} \sin \omega_0 t}{2(\omega_0^2+\lambda^2)^2}  \Big].  \nonumber
\label{gammafunc}
\end{align}
Defining the Rabi frequency of the laser as
$
\Omega = -\frac{ g_0 \alpha_0}{ \pi } \sqrt{ \frac{\omega_0}{2\omega}} 
$
%
and the spontaneous decay rate \cite{grimm00} as 
$
\Gamma_{\mbox{\tiny s}} = \frac{V  |g_0|^2 \omega_0 \omega}{8 \pi^2 c^3} .
\label{gammasp}
$
%
Therefore, we can have the following quantity
$
\frac{\Gamma_s |\Omega|^2}{\Delta^2}= \frac{V \omega_0^2 |\alpha_0|^2 }{16 \pi^4 c^3 \Delta^2} |g_0|^4.
$
%
Averaging this value over the solid angle $dO$ we get 
$
\frac{\Gamma_s |\Omega|^2}{\Delta^2}= \frac{ \omega_0^4 |\alpha_0|^2 |d|^4 }{5 \pi^3 c^3 \hbar^2 \epsilon_0^2 V\Delta^2}
$
Using this expression in the above equation 
we obtain the decoherence function.
Now averaging the function $\Gamma_s$ over the solid angle and using the result
$
\langle |g_0|^2 \rangle = \frac{8 \pi \omega_0 |d|^2}{3 \hbar \epsilon_0 V}
$
%
we get the standard result 
$
\Gamma_{\mbox{\tiny s}} = \frac{\omega_0^3  |d|^2}{3 \pi \hbar \epsilon_0 c^3}
$ \cite{grimm00},
%
where we have set $ \omega = \omega_0 $.


\end{document}